# Tunable Metamaterials Fabricated by Fiber Drawing

SIMON FLEMING,[1,*] ALESSIO STEFANI,[1,2] XIAOLI TANG,[1] ALEXANDER ARGYROS,[1] DANIEL KEMSLEY,[1] JAMES CORDI,[1] AND RICHARD LWIN[1]

[1]*Institute of Photonics and Optical Sciences (IPOS), School of Physics, The University of Sydney, NSW, 2006, Australia*
[2]*DTU Fotonik, Department of Photonics Engineering, Technical University of Denmark, DK-2800 Kgs. Lyngby, Denmark*
*Corresponding author: simon.fleming@sydney.edu.au

We demonstrate a practical scalable approach to the fabrication of tunable metamaterials. Designed for THz wavelengths, the metamaterial is comprised of polyurethane filled with an array of indium wires using the well-established fiber drawing technique. Modification of the dimensions of the metamaterial provides tunability: by compressing the metamaterial we demonstrated a 50% plasma frequency shift using THz time domain spectroscopy. Releasing the compression allowed the metamaterial to return to its original dimensions and plasma frequency, demonstrating dynamic reversible tunability.

**OCIS codes:** (160.3918) Metamaterials; (160.1245) Artificially engineered materials; (220.4000) Microstructure fabrication.

## 1. INTRODUCTION

In recent years metamaterials have been the subject of substantial research interest due to their extraordinary properties. These artificial composite materials, consisting typically of precise, sub-wavelength scaled, periodic metal features embedded in a dielectric, can be used to realize devices with a range of exotic functions from sub-diffraction imaging to cloaking [1-6]. Their properties derive primarily from their structure or geometry, rather than their constituent materials. Whilst fabrication of such structures of dissimilar materials on sub-wavelength scales is challenging, there have been numerous demonstrations [7], although few are scalable [8-10]. We have successfully fabricated metamaterials in volume by combining two scalable and well-established techniques: fiber drawing and the Taylor wire process [10-12].

It would be highly desirable to be able to dynamically tune the properties of a metamaterial. Within the many applications of tunable metamaterials, the underlying purpose has often been to adjust and move areas of high and low absorption, which is typically due to a resonance [13-15]. Many designs and material combinations have been used to realize tunable metamaterials operating in a wide range of frequencies [13-15]. Despite the introduction of more and more solutions, two major challenges remain: tunable metamaterials for high frequencies and metamaterials with a large tuning range. In fact, to date, most tunable metamaterials operate in the microwave frequencies [14]. Complexity in the structures that allow tunability is one of the aspects that make them difficult to scale for higher frequencies. As for how much the properties can be modified, tunability of around 30% was reported in the THz region [16-17] and up to 50% in the near-IR [18].

In this article, we report the proof of concept demonstration of tunable metamaterials, fabricated by fiber drawing, using polymers with relatively low Young's modulus, permitting tuning of metamaterial properties through reversible geometric deformation. A wire array is used to show that in this case a material change leads to wide tunability at the THz frequencies despite the simple structure.

## 2. TUNABLE METAMATERIALS BASED ON STRUCTURAL DEFORMATION

Our fiber drawing approach to fabricating metamaterials can be applied in the THz spectral region using polymers. The start is a macroscopic preform, made by any of the techniques for fabricating microstructured fiber preforms, such as stacking and drawing. The holes in the preform are filled with a metal with a sufficiently low melting temperature that it is molten at the polymer drawing temperature. The structure is then drawn such that the polymer is still highly viscous and the liquid metal takes the form of the holes which reduce in diameter during drawing [11-12].

Figure 1 shows a schematic of a typical wire array metamaterial considered here. It comprises an array of metal wires with diameter $d$, separated by distance $a$, embedded in a dielectric of refractive index $n_d$. The permittivity of this material to electromagnetic radiation polarized parallel to the wires can be described [19] by a Drude model where the plasma frequency, $f_p$, depends almost entirely on the geometry, not the materials:

$$f_p^2 = \frac{(c/n_d)^2}{2\pi A_{cell} \ln\left[\frac{a^2}{d(2a-d)}\right]} \quad (1)$$

where $c$ is the speed of light, and $A_{cell}$ is the area of the unit cell.

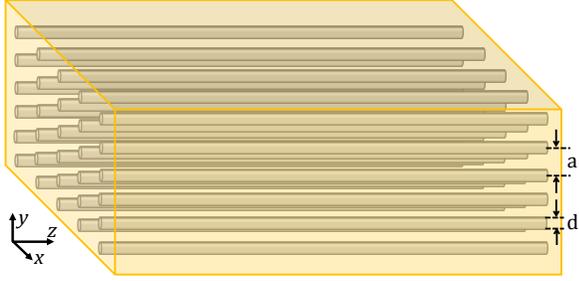

Fig. 1. Schematic of wire array metamaterial (wires are shown grey, dielectric pale amber).

The conceptually simplest approach to realizing a tunable metamaterial with this fabrication method would be a structure that possessed both periodic arrays of wires and air holes (Fig. 2a), and where the material had a sufficiently low Young's modulus that the polymer bridges between the air holes could be stretched or compressed, allowing the whole structure to expand or shrink in the transverse plane (Fig. 2b). The spacing, unit cell and "average" dielectric index would change, and the plasma frequency and hence permittivity could be tuned by compression or expansion.

To date we have used a range of polymers with relatively high Young's moduli, such as polymethylmethacrylate (PMMA) (Young's modulus 2.4–3.4 GPa [20]), which will not permit any significant stretching of this type. However, compression of the whole structure in one dimension, essentially by bending of the polymer bridges seemed feasible (Fig. 2c).

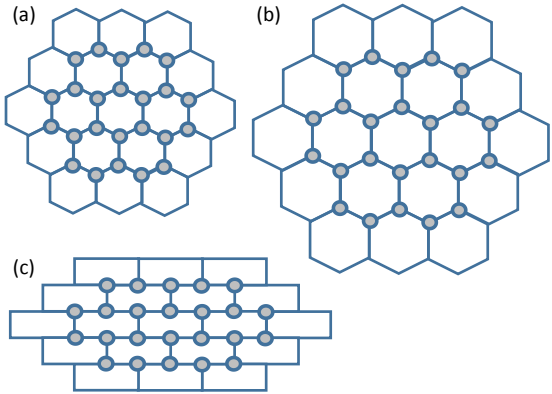

Fig. 2. Approaches to tunability through deformation (wires are shown in grey, polymer blue, air white). (a) As made, (b) uniform expansion, (c) unidirectional (vertical) compression.

The analysis of the effect of this type of deformation on the metamaterial properties is rather more complicated [21] as the separation, $a$, is generally no longer single valued. However if the difference between these values is less than ~50%, which is larger than we anticipate experimentally achieving, Eq. 1 is a good approximation when used with the average value for $a$.

We fabricated a metamaterial fiber of similar design with periodic wires and air holes (Fig 3a), and it readily deformed, although not reversibly, effectively crushing (Fig 3b). Whilst this approach may be useful for post-fabrication tuning, it is not the dynamic tunability performance we were seeking.

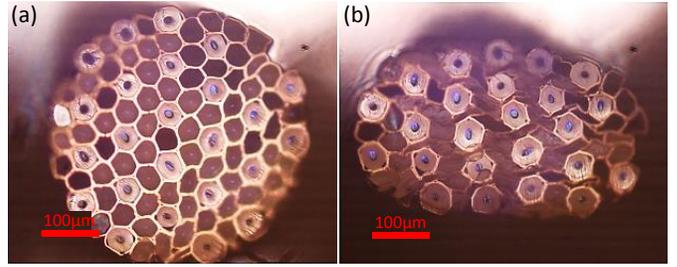

Fig. 3. Fabricated metamaterial based on PMMA and bending rather than compression: (a) as fabricated, (b) deformed/tuned.

Clearly a lower Young's modulus polymer was required and we selected polyurethane, with Young's modulus 2-3 MPa [22]. This is sufficiently low that a design without air holes, relying solely on compression of the polymer, along the lines of Fig 2c, is feasible, and simpler to fabricate as an initial demonstration.

The polyurethane based tunable metamaterial was fabricated using the capillary stacking method commonly used to manufacture preforms for microstructured optical fibers and drawn metamaterials [11-12]. It consisted of an array of indium wires arranged in an hexagonal lattice, with each wire encased in a thin PMMA tube and a thick polyurethane tube (Fig. 4). A thin PMMA tube was used to prevent distortion of the indium wires during the drawing process and when the metamaterial was compressed during operation. The role of the polyurethane was purely to provide a bulk material that provides high, and reversible, tunability through compression. An outer jacket of PMMA was used to contain the wire array stack and provide rigidity during the drawing process, but was removed prior to measurements.

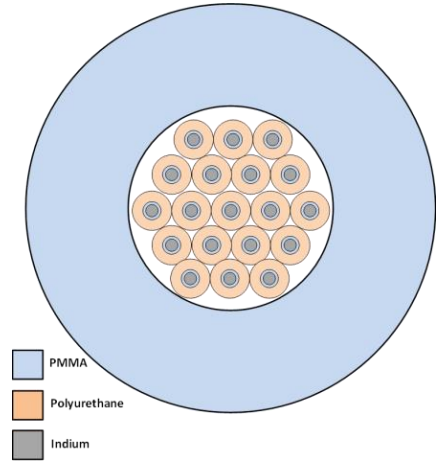

Fig. 4. Preform design used to fabricate polyurethane based tunable metamaterial.

The indium filled PMMA/polyurethane wire was then drawn down using the draw tower to a size appropriate for stacking into the outer PMMA jacket, using a feed speed of 5 mm/min, draw speed of 35 mm/min and furnace temperature of 178°C, to achieve an OD of 1mm. These indium filled fibers were then cut using a razor blade to lengths of around 300 mm and stacked into the outer PMMA jacket (ID 6 mm/OD 8 mm) to create the resulting preform (Fig. 5a), and annealed at 90°C for 24 hours to remove any residual stress.

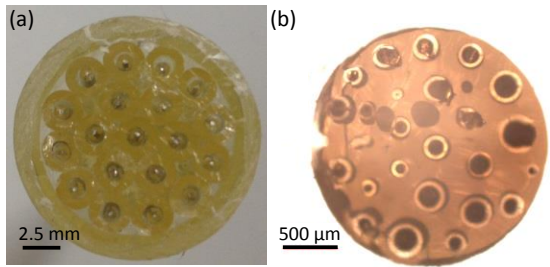

Fig. 5. Polyurethane metamaterial fabrication: (a) stacked preform and (b) final drawn structure.

Each indium/PMMA/polyurethane element was made by first sleeving the PMMA tube with an inner diameter (ID) 2.5 mm and outer diameter (OD) 3.15 mm into a polyurethane tube (ID 3.18 mm/OD 6.35 mm) in the polymer fiber draw tower, using a feed speed of 10mm/min and furnace temperature of 168°C. Vacuum was applied between the PMMA and polyurethane tubes during the drawing to cause the two surfaces to adhere. The final dimension of the sleeved tube was ID 1 mm/OD 2.5 mm.

Indium was introduced into the sleeved tube by first melting the metal and drawing it into the tube using a syringe. Approximately 200-300 mm length of indium was drawn into the sleeved tube before solidification caused the process to terminate.

The stacked preform was drawn in a two stage process, first at feed speed 5 mm/min, draw speed 20 mm/min and furnace temperature of 180°C to make a 600 mm length of intermediate cane. This was then drawn again at feed speed 10 mm/min, draw speed 40 mm/min and furnace temperature of 180°C to make the final fiber of OD 2 mm and length 1200 mm (Fig. 5b). The indium wires were around 90 μm diameter and spaced 290 μm apart, creating a metamaterial wire array with a plasma frequency of 0.37 THz calculated using Eq. 1 and assuming an hexagonal lattice. The resulting cross-section shows variations in wire sizes and wire positions deviating from the original design. Both factors are related to the different drawing conditions required for PU, PMMA and indium. Despite the variations in wire sizes in the cross-section, longitudinally the wires are constant in diameter for lengths over the 10 cm required.

To make the metamaterial tunable, the outer PMMA jacket was etched away (over 4 hours) using acetone, leaving the polyurethane as the bulk outer material of OD 1.5 mm. As shown in Fig. 6, the polyurethane based metamaterial can be easily compressed (using tweezers) and returns to its original shape when released.

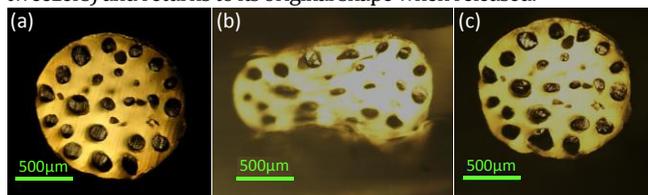

Fig. 6. Changing the geometry of the metamaterial using tweezers. (a) Original, (b) compressed, (c) released after compression.

The sample shown in Fig. 6 has some fluctuations in wire diameters and spacing. Various samples were fabricated with different drawing parameters. At this time the ideal condition has not been found yet, but it was possible to realize more regular and smaller structures. As an example, a different structure is shown in Fig. 7. This structures has a larger number of wires, the wire size is about 20 μm and the spacing 180 μm. Despite it not being circular, the wire size and spacing are more uniform. The calculated plasma frequency for this sample is 0.386 THz.

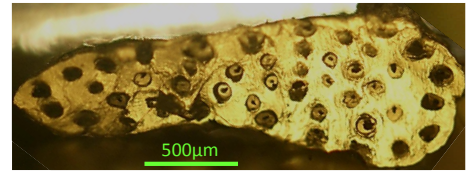

Fig. 7. Fabricated metamaterial with improved wire geometry.

## 3. TUNABILITY MEASUREMENTS

A linear array of metamaterial fibers was required for the THz measurement. 10 mm lengths of fibers were cut using a razor blade and arranged into a 10 mm x 10 mm linear array (as depicted from Fig. 8a). Adhesive tape was used to join the fibers together. This provided sufficient flexibility during compression of the metamaterial array, and was transparent to THz radiation. To demonstrate the tunability due to compression, the metamaterial array was sandwiched between two Zeonex® plates of 5 mm thickness, which are transparent in the THz. The amount of compression was controlled by tightening screws connecting the two Zeonex plates, thus compressing the metamaterial array. A schematic of the measurement configuration is shown in Fig. 8b.

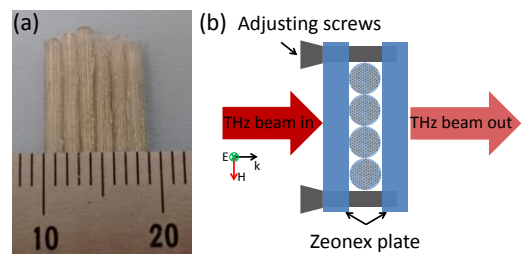

Fig. 8. (a) The linear array of metamaterial fibers. (b) Schematic of the measurement configuration.

The fabricated samples were characterized by using THz time domain spectroscopy (TDS). The transmission through the metamaterial fibers was measured and it was normalized to that of the THz system through the Zeonex plates without the metamaterial sample. The electric field of the THz beam was aligned along the direction of the metal wires to excite the plasma frequency.

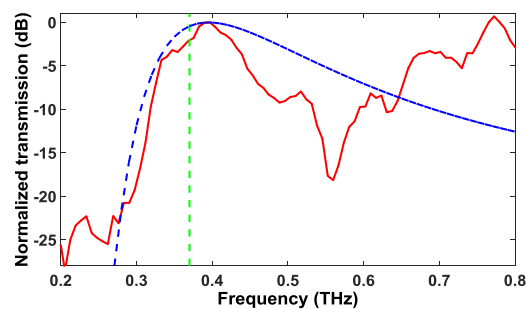

Fig. 9. Normalized transmission (red line) through the uncompressed metamaterial, with simple Drude model (blue line) best fit using plasma frequency of 0.395 THz, and showing the theoretical plasma frequency of 0.37 THz (green line) of the nominal structure.

The transmission in the uncompressed sample was first measured with the screws loose. This first measurement confirmed that the fabricated structure had the expected plasma frequency. Three different fabricated samples were measured and, as expected, they all showed similar responses. To show the working principle, only the results for the sample of Fig. 6 are reported.

Figure 9 shows the normalized transmission of the uncompressed sample. The transmission measurements were renormalized so that 0 dB corresponds to the maximum of transmission. A transition between low and high transmission is clearly visible, indicating the transition between metallic and dielectric behavior. In all measurements, the maximum signal is at least 20 dB above the noise level. The measured plasma frequency was estimated as 0.395 THz by fitting a simplified Drude model, as shown by the blue dashed line in Fig. 9. This value for the plasma frequency is in good agreement with the expected value of 0.37 THz (green vertical line in Fig. 9) despite the variation in wire diameters. Other than the transition from low to high transmission, the spectrum shows various spectral features known for fiber metamaterials [10] which are not relevant in order to demonstrate tunability.

The sample was then compressed an arbitrary amount by tightening the screws and the transmission was recorded. After that, the structure was compressed further to the maximum compression achievable with the setup and again the transmission was recorded. The results of the compression experiments are shown in the top panel of Fig. 10. The plasma frequency shifts to higher frequencies with compression, as expected, with a maximum shift of 0.2 THz, which corresponds to a tunability of 50%. Using Eq. 1, and assuming only the spacing between the wires is reduced, a 21% compression of the structure (separation reduced from 290 μm to 227 μm) produces the desired shift for the structure used in the measurements of Fig. 10, Similarly, the same shift can be obtained with a 28% compression for the structure shown in Fig. 7 (from 180 μm to 130 μm).

To ensure the shift was not a permanent effect due to the structure crushing, the pressure from the screws was released and the transmission recorded. The central panel of Fig. 10 shows the full recovery of the sample which returns to the original spacing and thus, plasma frequency. Small alterations in the spectrum from the original arise from the fibers' positions relative to each other settling the first time they are compressed. The sample was than compressed again and released a second time to make sure the results were consistent. As can be seen from Fig. 10 the plasma frequency can be tuned to its maximum value and recovers each time.

## 4. CONCLUSIONS

We have successfully fabricated a fiber based tunable metamaterial using polyurethane and indium, which operates in the THz frequency region through mechanical deformation of the geometry of the structure of the metamaterial. We demonstrated a positive plasma frequency shift of 50% by compressing the metamaterial array, returning to the original plasma frequency upon release.

**Funding Information.** The Australian Research Council under the Discovery Project scheme numbers DP 140104116 and 170103537 (SF); the Eugen Lommel Stipend and Marie Sklodowska-Curie grant of the European Union's Horizon 2020 research and innovation programme (708860) (AS); Postdoctoral International Exchange Program jointly sponsored by China Postdoctoral Science Foundation and The University of Sydney (XT).

**Acknowledgment**. We thank Boris Kuhlmey and Scott Brownless for useful discussions. This work was performed at the OptoFab Node of The Australian National Fabrication Facility, utilizing NCRIS and NSW State Government funding.

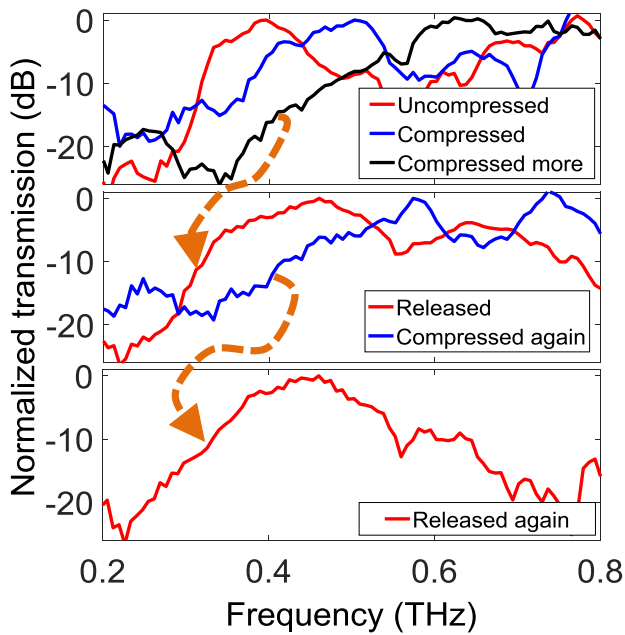

Fig. 10. Transmission through the metamaterial sample during the compression and release cycles.